\documentclass[12pt]{article}
\usepackage{amsmath, amssymb}
\begin{document}
\date{}
\title{An $su(1,1)$ algebraic approach for the relativistic Kepler-Coulomb problem}

\author{M. Salazar-Ram\'{\i}rez$^{a}$, D. Mart\'{\i}nez$^{b}$,\footnote{{\it E-mail address:} dmartinezs77@yahoo.com.mx}\\ R. D. Mota$^{c}$ and V. D. Granados$^{a}$} \maketitle

\begin{minipage}{0.9\textwidth}

\small $^{a}$ Escuela Superior de F\'{\i}sica y Matem\'aticas,
Instituto Polit\'ecnico Nacional,
Ed. 9, Unidad Profesional Adolfo L\'opez Mateos, 07738 M\'exico D F, M\'exico.\\

\small $^{b}$ Universidad Aut\'onoma de la Ciudad de M\'exico,
Plantel Cuautepec, Av. La Corona 320, Col. Loma la Palma,
Delegaci\'on
Gustavo A. Madero, 07160, M\'exico D. F., M\'exico.\\

\small $^{c}$ Unidad Profesional Interdisciplinaria de
Ingenier\'{\i}a y Tecnolog\'{\i}as Avanzadas, IPN. Av. Instituto
Polit\'ecnico Nacional 2580, Col. La Laguna Ticom\'an, Delegaci\'on
Gustavo A.
Madero, 07340 M\'exico D. F., M\'exico.\\

\end{minipage}

\begin{abstract}
We apply the Schr\"odinger factorization method to the radial
second-order equation for the relativistic Kepler-Coulomb problem.
From these operators we construct two sets of one-variable radial
operators which are realizations for the $su(1,1)$ Lie algebra. We
use this  algebraic structure to obtain the energy spectrum and
the supersymmetric ground state for this system.
\end{abstract}
{\it PACS:} 03.65.Pm; 03.65.Fd; 03.65.Ge; 02.20.Sv\\
{\it Keywords:} Relativistic hydrogen atom; Factorization methods; $su(1,1)$ Lie algebra\\

\section{Introduction}
Compact and non-compact symmetries play a central role to study
many properties of quantum systems because they form the basis for
selection rules that forbid the existence of certain states and
processes. Also, from a given solution, symmetries allow to obtain
new solutions of the dynamical equations. Moreover, the conserved
quantity associated with a given symmetry represents an
integrability condition.

Generators of compact and non-compact algebras for a given
Hamiltonian have not been constructed by a systematic method but
intuitively found and forced to close an algebra
\cite{WYBOURNE,BARUT,ENGLEFIELD}. In particular, the $su(1,1)\sim
so(2,1)$ non-compact algebra is the smallest dynamical symmetry
admitting infinite-dimensional representations.  This symmetry has
been successfully applied to formulate algebraic approaches to
many non-relativistic quantum problems
\cite{MILLER,AL1,AL2,AL3,FRANK1,FRANK2,FRANK3,WU,KERIMOV,ENGLE,LEVAI},
where the corresponding realizations were given in terms of two
variables.  Also, some one-variable realizations for the $su(1,1)$ algebra
have been introduced \cite{WYBOURNE}. Recently, it has been shown
that the Schr\"odinger factorization operators can be used to
construct the $su(1,1)$ algebra generators for two- \cite{DANIEL}
and $N$-dimensional systems \cite{NDIM}.

The relativistic Kepler-Coulomb problem is one of the few exactly
solvable potentials in physics and it has been studied in several
ways: analytical \cite{BJOR,THALL1,THALL2,GREINER,ROBIN1,ROBIN2},
factorization (algebraic) methods  \cite{IH,CHINOS1,CHINOS2},
shape-invariance \cite{LIMA} and  SUSY QM for the first-
\cite{THALL1,SUKU} and second-order equations \cite{JARVIS}. The
solubility of this problem is due to the conservation of the total
angular momentum, the Dirac and Lippmann-Johnson operators
\cite{DAHL}. The first two ones are due to the existence of spin,
and the later explains the degeneracy in the eigenvalues of the
Dirac operator of the energy spectrum  and  it reduces to the
Runge-Lenz vector in the non-relativistic limit. Symmetries and
SUSY QM are not independent. Indeed, it has been shown that
supersymmetry is generated by the Lippmann-Johnson operator
\cite{DAHL}.

In a series of papers some realizations of compact and non-compact
Lie algebras for the uncoupled second-order radial equations
corresponding to the relativistic Kepler-Coulomb problem have been
introduced \cite{MEX1,MEX2,MEX3}. However, in all these works the
origin of the generators was not given. Moreover, in order to
close the $su(1,1)$ Lie algebra these generators were forced to
depend on an extra variable which plays the role of a phase.

The aim of this paper is to study the relativistic Kepler-Coulomb problem from an $su(1,1)$ algebraic approach by using the Schr\"odinger factorization and without the introduction of any additional variable.

In section 2, we obtain the uncoupled second-order differential equations satisfied by the radial components. In section 3, we obtain the $su(1,1)$ algebra generators by applying the Schr\"odinger factorization to an uncoupled second-order differential equation. These results allow us to obtain the energy spectrum for this system. In section 4, the SUSY ground state and the action of the $su(1,1)$ algebra generators on the radial eigenstates are found. Finally, we give some concluding remarks.

\section{The relativistic Kepler-Coulomb radial equation}
The Dirac radial equation for the Kepler-Coulomb problem is
\cite{BJOR,SUKU}
\begin{equation}
\begin{pmatrix}
\frac{dG^{(1)}_{k}}{dr}\\
\frac{dG^{(2)}_{k}}{dr}
\end{pmatrix}+
\frac{1}{r}\begin{pmatrix}
k & -\gamma \\
\gamma & -k \end{pmatrix}
\begin{pmatrix}
G^{(1)}_{k}\\
G^{(2)}_{k}
\end{pmatrix}=\begin{pmatrix}
0 & \alpha_1\\
\alpha_2 & 0 \end{pmatrix}\begin{pmatrix}
G^{(1)}_{k}\\
G^{(2)}_{k}
\end{pmatrix},\label{dirac1}
\end{equation}
where $k$ is the eigenvalue of the Dirac operator
$K=-\left(\vec{\sigma}\cdot\vec{L}+\mathbf{1}\right),\gamma=\frac{ze^2}{c\hbar},
\alpha_1=m+E,\alpha_2=m-E, \left|k\right|=j+\frac{1}{2},
k=\pm1,\pm2,\pm3,...,
\vec{\sigma}=\left(\sigma_1,\sigma_2,\sigma_3\right)$ are the
Pauli matrices, $\vec{L}$ is the angular momentum operator and
$\mathbf{1}$ is the $2\times 2$ unit matrix.

By introducing the new variable $\rho = Er$ equation
(\ref{dirac1}) leads to the pair of equations
\begin{align}\label{acop1}
\left(\frac{k}{s}+\frac{m}{E}\right)F^{(2)}_{k}&=\left(\frac{d}{d\rho}+\frac{s}{\rho}-\frac{\gamma}{s}\right)F^{(1)}_{k},\\\label{acopl2}
\left(\frac{k}{s}-\frac{m}{E}\right)F^{(1)}_{k}&=\left(-\frac{d}{d\rho}+\frac{s}{\rho}-\frac{\gamma}{s}\right)F^{(2)}_{k},
\end{align}
where
\begin{equation}
\begin{pmatrix}
F^{(1)}_{k}\\
F^{(2)}_{k}
\end{pmatrix}=D\begin{pmatrix}
G^{(1)}_{k}\\
G^{(2)}_{k}
\end{pmatrix}\equiv \begin{pmatrix}
k+s&-\gamma\\
-\gamma & k+s\end{pmatrix}\begin{pmatrix}
G^{(1)}_{k}\\
G^{(2)}_{k}
\end{pmatrix}
\end{equation}
and
\begin{equation}
s=\sqrt{k^2-\gamma^2}\label{s}.
\end{equation}

From these relations we obtain the uncoupled second-order
differential equations
\begin{align}
\left(-\frac{d^2}{d\rho^2} +
\frac{s\left(s\pm1\right)}{\rho^2}+\frac{\gamma^2}{s^2}-\frac{2\gamma}{\rho}\right)F^{(1,2)}_{k}&=
\left(\frac{k}{s}+\frac{m}{E}\right)\left(\frac{k}{s}-\frac{m}{E}\right)F^{(1,2)}_{k},\label{desacop1}
\end{align}
where the superscript $(1,2)$ corresponds to $(+,-)$ in the
centrifugal term, respectively. Equation (\ref{s}) allows us to
rewrite equation (\ref{desacop1}) as
\begin{align}
\left(-\rho^2\frac{d^2}{d\rho^2}+\xi^2\rho^2-2\gamma\rho\right)F^{(1)}_{k}& =-s\left(s+1\right)F^{(1)}_{k},\label{desajun1}\\
\left(-\rho^2\frac{d^2}{d\rho^2}+\xi^2\rho^2-2\gamma\rho\right)F^{(2)}_{k}& =-s\left(s-1\right)F^{(2)}_{k},\label{desajun2}
\end{align}
where
\begin{equation}
\xi^2=\frac{m^2}{E^2}-1.\label{xi}
\end{equation}
It must be emphasized that equation (\ref{desajun1}) is formally
obtained from equation (\ref{desajun2}) by making $s\rightarrow
s+1$. Hence, by defining $\psi_s\equiv F^{(2)}_{k}$, we conclude
that
\begin{equation}
F^{(1)}_{k}\propto\psi_{s+1}.
\end{equation}
Thus, the solution for the Dirac equation in spinorial form is
\begin{equation}
\Phi_{k}\equiv\begin{pmatrix}
F^{(1)}_{k}\\
F^{(2)}_{k}
\end{pmatrix}=
\begin{pmatrix}
\psi_{s+1}\\
\psi_{s}
\end{pmatrix}.\label{spinor}
\end{equation}

\section{The Schr\"odinger Factorization and the $su(1,1)$ Lie algebra}
In order to factorize the left-hand side operator of equations
(\ref{desajun2}), we apply the Schr\"odinger factorization
\cite{SCH1,DANIEL1}. Thus, we propose a pair of first-order
differential operators such that
\begin{equation}
\left(\rho\frac{d}{d\rho}+a\rho+b\right)\left(-\rho\frac{d}{d\rho}+c\rho+f\right)\psi_{s}=g\psi_{s},
\label{sch}
\end{equation}
where $a$, $b$, $c$, $f$ and $g$ are constants to be determined.
Expanding this expression and comparing it with equation
(\ref{desajun2}) we obtain
\begin{align}
a=c=\pm\xi,\hspace{2ex}f=1+b=\mp\frac{\gamma}{\xi},\hspace{2ex} g=b(b+1)-s(s-1).
\end{align}
Using these results, equation (\ref{desajun2}) is equivalent to
\begin{align}
\left(B_\mp\mp1\right)B_\pm\psi_{s} =\left[\left(\frac{\gamma}{\xi}\pm\frac{1}{2}\right)^2-\left(s-\frac{1}{2}\right)^2\right]\psi_{s},\label{Facdesc1}
\end{align}
where
\begin{equation}\label{OpeFac}
B_\pm =\mp\rho\frac{d}{d\rho} +\xi\rho-\frac{\gamma}{\xi}.
\end{equation}
From equation (\ref{desajun2}) we obtain the operator
\begin{align}
\Sigma_3\psi_{s}\equiv\frac{1}{2\xi}\left(-\rho\frac{d^2}{d\rho^2}
+\xi^2\rho+\frac{s\left(s-1\right)}{\rho}\right)\psi_{s}=\frac{\gamma}{\xi}\psi_{s}.\label{T3dif2or1}
\end{align}
Therefore, using equation (\ref{OpeFac}) we define a new pair of
operators
\begin{align}
\Sigma_\pm & \equiv\mp\rho\frac{d}{d\rho}+\xi\rho-\Sigma_3.\label{Tmasmen1}
\end{align}

In order to establish the properties of the operators $\Sigma_3$
and $\Sigma_\pm$ we define the inner product on the Hilbert space
spanned by the radial eigenfunctions of the relativistic
Kepler-Coulomb problem as \cite{ADAMS}
\begin{equation}
\left(\phi,\zeta\right)\equiv\int_0^\infty \phi^{*}\left(\rho\right)\zeta\left(\rho\right)\rho^{-1}d\rho.\label{ortonor}
\end{equation}
Thus, it follows that the operator $\Sigma_3$ is hermitian.
Moreover, using equations (\ref{Tmasmen1}) and (\ref{ortonor}) we
prove that the operators $\Sigma_\pm$ are hermitian conjugates,
\begin{align}
\Sigma_\pm& =\Sigma_\mp^\dagger.
\end{align}

By direct calculation it is immediate to show that the operators
$\Sigma_\pm$ and $\Sigma_3$ close the $su(1,1)$ Lie algebra
\begin{align}
\left[\Sigma_\pm,\Sigma_3\right]& =\mp\Sigma_\pm,\label{Relcom11}\\
\left[\Sigma_+,\Sigma_-\right]& =-2\Sigma_3,\label{Relcom21}
\end{align}
with quadratic Casimir operator
\begin{align}
\Sigma^2=-\Sigma_{+}\Sigma_{-}+\Sigma_3^2 - \Sigma_3.\label{Cas1}
\end{align}
From equations (\ref{T3dif2or1}) and (\ref{Tmasmen1}) we find that
the eigenvalue equation for this operator is
\begin{align}
\Sigma^2\psi_{s}=s\left(s-1\right)\psi_{s}.\label{CasEn1}
\end{align}

The theory of unitary irreducible representations of the $su(1,
1)$ Lie algebra has been studied in several works \cite{ADAMS,
ADAMS1} and it is based on the equations
\begin{align}
C^2\vert\mu\:\nu\rangle& =\mu(\mu+1)\vert\mu\:\nu\rangle,\label{c2}\\
C_3\vert\mu\:\nu\rangle& =\nu\vert\mu\:\nu\rangle\label{c3},\\
C_\pm\vert\mu\:\nu\rangle& =\left[(\nu\mp\mu)(\nu\pm\mu\pm1)\right]^{1/2}\vert\mu\:\nu\pm1\rangle\label{C+-},
\end{align}
where $C^2$ is the quadratic Casimir operator, $\nu=\mu+q+1$,
$q=0,1,2,...$ and $\mu>-1$. The last relation means that
$C_+$($C_-$) are the raising (lowering) operators for $\nu$.

Therefore, from equations (\ref{CasEn1}) and (\ref{c2}), and
(\ref{T3dif2or1}) and (\ref{c3}), we find
\begin{align}
\mu_{s}& =s-1,\label{mu1}\\
\nu_{s}& =n_{s}+s=\frac{\gamma}{\xi},\label{nu1}
\end{align}
respectively, with $n_{s}=0,1,2,...$.

If we consider that the operators $\Sigma_\pm$ leave fixed the
quantum number $\mu_s$ then, from equation (\ref{mu1}), the values
of $s$ remain fixed. This is because the change
$\nu_s\rightarrow\nu_s\pm1$ induced by the operators $\Sigma_\pm$
on the basis vectors $\vert\mu_s\:\nu_s\rangle$ implies
$n_s\rightarrow n_s\pm1$. Thus, by setting $\vert\mu_s\:\nu_s\rangle\rightarrow\psi_s^{n_s}$
and from equations (\ref{C+-}), (\ref{mu1}) and (\ref{nu1}) we find
\begin{align}
\Sigma_{\pm}\psi_{s}^{n_s}=Q_{\pm}^{\left(n_s,s\right)}\psi_{s}^{n_s\pm1},\label{Sigma+-}
\end{align}
with $Q_{\pm}^{\left(n_s,s\right)}=\left[\left(n_s+s+1\mp s\right)\left(n_s+s+1\pm s\pm1\right)\right]^{1/2}$ a real number.

Using equations (\ref{xi}) and (\ref{nu1}) we find that the energy
spectrum for the lower component of the spinor given in equation 
(\ref{spinor}), $\psi_{s}^{n_{s}}$, is
\begin{equation}
E_{s}=m\left[1+\frac{\gamma^2}{(n_{s}+s)^2}\right]^{-1/2}.\label{E1}
\end{equation}

As was emphasized above, by performing the change $s\rightarrow
s+1$ to $\psi_s^{n_s}$ we obtain the upper component of the spinor
$\Phi_k$,  $\psi_{s+1}^{n_{s+1}}$. In this way, its corresponding
differential operators are
\begin{align}
\Xi_3&\equiv\frac{1}{2\xi}\left(-\rho\frac{d^2}{d\rho^2}
+\xi^2\rho+\frac{s\left(s+1\right)}{\rho}\right),\label{T3dif2or2}\\
\Xi_\pm & \equiv\mp\rho\frac{d}{d\rho}+\xi\rho-\Xi_3\label{Tmasmen2}
\end{align}
and by direct calculation we show they satisfy the $su(1,1)$ Lie algebra
\begin{align}
\left[\Xi_\pm,\Xi_3\right]& =\mp\Xi_\pm,\label{Relcom12}\\
\left[\Xi_+,\Xi_-\right]& =-2\Xi_3.\label{Relcom22}
\end{align}
From equation (\ref{Sigma+-}) we obtain that the action of the
ladder operators $\Xi_\pm$ on the functions $\psi_{s+1}^{n_{s+1}}$
is
\begin{align}
\Xi_{\pm}\psi_{s+1}^{n_{s+1}}=Q_{\pm}^{\left(n_{s+1},s+1\right)}\psi_{s+1}^{n_{s+1}\pm1}.\label{Xi+-}
\end{align}
Hence, from equation (\ref{E1}) we find that the energy spectrum
for the function $\psi_{s+1}^{n_{s+1}}$ is
\begin{equation}
E_{s+1}=m\left[1+\frac{\gamma^2}{(n_{s+1}+s+1)^2}\right]^{-1/2}.\label{E2}
\end{equation}
Since $\psi_{s+1}^{n_{s+1}}$ and $\psi_{s}^{n_{s}}$ are the
components for the spinor $\Phi_{k}$, they must have the same
energy. This means that $E_s=E_{s+1}$. Therefore, from equations
(\ref{E1}) and (\ref{E2}) we obtain
\begin{equation}
n\equiv n_{s}=n_{s+1}+1,\label{n}
\end{equation}
where $n=0,1,2,3,...$ is the radial quantum number. Thus, the
energy spectrum for the relativistic Kepler-Coulomb problem is
\begin{equation}
E=m\left[1+\frac{\gamma^2}{(n+s)^2}\right]^{-1/2},
\end{equation}
where the positiveness of $E$ ensures that the components of the
spinor given in (\ref{spinor}) are quadratically integrable
\cite{MEX1}. In this way, $\Phi_k^n$ is given by
\begin{equation}
\Phi_{k}^n=\begin{pmatrix}
F_{n\:k}^{(1)}\\
F_{n\:k}^{(2)}
\end{pmatrix}=
\begin{pmatrix}
\psi_{s+1}^{n-1}\\
\psi_s^n
\end{pmatrix}.\label{spinor1}
\end{equation}
This result, which has been obtained from the theory of unitary representations, can be deduced from an analytical approach as it is shown in the next section.

\section{Schr\"odinger and SUSY QM ground states}
For $n=0$ and from equations (\ref{Sigma+-}) and (\ref{n}) we find
that only the state
\begin{equation}
\psi_s^{0}=N\rho^s e^{-\gamma\rho/s},
\end{equation}
is normalizable respect to the inner product defined in expression
(\ref{ortonor}) and  satisfies the differential equation
$\Sigma_-\psi_s^{0}=0$. Also, for $n=0$ and from equation (\ref{n})
we find $n_{s+1}=-1$. For these values, $Q_{-}^{\left(-1,s+1\right)}$ results to be a complex number.
Thus, from the theory of unitary representations and equation (\ref{Xi+-}), the function
$\psi_{s+1}^{-1}$ is non-normalizable \cite{ADAMS1}. Since this
function is not a physically acceptable solution, the spinor corresponding to $n=0$ is
\begin{equation}
\Phi_{k}^0=\begin{pmatrix}
0\\
\rho^s e^{-\gamma\rho/s}
\end{pmatrix},\label{spinor2}
\end{equation}
whom we denote as the Schr\"odinger ground state. Notice that $\Phi_k^0$ is equal to the SUSY ground state for the
relativistic Kepler-Coulomb problem found in \cite{SUKU,DAHL}.

From an analytical approach, to determine $\Phi_k^n$ for
higher levels we solve the differential equation (\ref{desajun2})
by proposing
\begin{equation}
F^{(2)}_{k}=\rho^{s}e^{-\xi\rho} f(\rho).\label{Ff}
\end{equation}
Thus, $f(\rho)$ must satisfy
\begin{equation}
\left[y\frac{d^2}{dy^2}+(2s-y)\frac{d}{dy}+\frac{\gamma}{\xi}-s\right] f\left(y/2\xi\right)=0,
\end{equation}
where $y=2\xi\rho$. The solutions for this equation is the
confluent hypergeometric function
$_1F_1\left(-n,2s;y\right)$. Thus, we get
\begin{equation}
F_{n\:k}^{(2)}=N_2\rho^se^{-\xi\rho}\;_1F_1\left(-n,2s;2\xi\rho\right).\label{F1}
\end{equation}
In a similar way, the solution for equation (\ref{desajun1}) is 
\begin{equation}
F_{n\:k}^{(1)}=N_1\rho^{s+1}e^{-\xi\rho}\;_1F_1\left(-n+1,2s+2;2\xi\rho\right).\label{F2}
\end{equation}
This equations are in agreement with our results obtained from the theory of
unitary representations, equation (\ref{spinor1}). Moreover, it is known that $_1F_1(0,b;z)=1$, whereas $_1F_1(a,b;z)$ diverges for $a>0$ .
Thus, for $n=0$
\begin{equation}
F^{(2)}_{0\:k}=N_s\rho^se^{-\xi\rho},
\end{equation}
while the function $F^{(1)}_{0\:k}$ is not square-integrable and it must be taken as the zero funtion. This result is in accordance with equation (\ref{spinor2}) which has been obtained from an algebraic approach.

\setlength{\unitlength}{0.5mm}
\begin{figure}
\begin{picture}(170,160)
\linethickness{0.5mm}
\put(50,7){\makebox(0,0){\scriptsize $k$}}
\put(50,20){\makebox(0,0){\scriptsize $N=1$}}
\put(50,80){\makebox(0,0){\scriptsize $N=2$}}
\put(50,110){\makebox(0,0){\scriptsize $N=3$}}
\put(50,130){\makebox(0,0){\scriptsize $N=4$}}
\put(50,145){\makebox(0,0){\scriptsize $N=5$}}
\multiput(60,20)(0,30){1}{\line(1,0){10}}
\multiput(60,80)(0,60){1}{\line(1,0){10}}
\multiput(60,110)(0,60){1}{\line(1,0){10}}
\multiput(60,130)(0,60){1}{\line(1,0){10}}
\multiput(60,145)(0,60){1}{\line(1,0){10}}
\multiput(75,80)(0,60){1}{\line(1,0){10}}
\multiput(75,110)(0,60){1}{\line(1,0){10}}
\multiput(75,130)(0,60){1}{\line(1,0){10}}
\multiput(75,145)(0,60){1}{\line(1,0){10}}
\put(65,7){\makebox(0,0){\scriptsize $-1$}}
\put(80,7){\makebox(0,0){\scriptsize $1$}}
\multiput(95,83)(0,60){1}{\line(1,0){10}}
\multiput(95,113)(0,60){1}{\line(1,0){10}}
\multiput(95,133)(0,60){1}{\line(1,0){10}}
\multiput(95,148)(0,60){1}{\line(1,0){10}}
\multiput(110,113)(0,60){1}{\line(1,0){10}}
\multiput(110,133)(0,60){1}{\line(1,0){10}}
\multiput(110,148)(0,60){1}{\line(1,0){10}}
\put(100,7){\makebox(0,0){\scriptsize $-2$}}
\put(115,7){\makebox(0,0){\scriptsize $2$}}
\multiput(130,116)(0,60){1}{\line(1,0){10}}
\multiput(130,136)(0,60){1}{\line(1,0){10}}
\multiput(130,151)(0,60){1}{\line(1,0){10}}
\multiput(145,136)(0,60){1}{\line(1,0){10}}
\multiput(145,151)(0,60){1}{\line(1,0){10}}
\put(135,7){\makebox(0,0){\scriptsize $-3$}}
\put(150,7){\makebox(0,0){\scriptsize $3$}}
\multiput(165,139)(0,60){1}{\line(1,0){10}}
\multiput(165,154)(0,60){1}{\line(1,0){10}}
\multiput(180,154)(0,60){1}{\line(1,0){10}}
\put(170,7){\makebox(0,0){\scriptsize $-4$}}
\put(185,7){\makebox(0,0){\scriptsize $4$}}
\multiput(200,157)(0,60){1}{\line(1,0){10}}
\put(205,7){\makebox(0,0){\scriptsize $-5$}}
\put(220,7){\makebox(0,0){\scriptsize $5$}}
\linethickness{0.2mm}
\put(100,135){\vector(0,1){10}}
\put(115,135){\vector(0,1){10}}
\put(96,140){\makebox(0,0){\scriptsize $\Sigma_+$}}
\put(111,140){\makebox(0,0){\scriptsize $\Xi_+$}}
\put(100,130){\vector(0,-1){14}}
\put(115,130){\vector(0,-1){14}}
\put(96,123){\makebox(0,0){\scriptsize $\Sigma_-$}}
\put(111,123){\makebox(0,0){\scriptsize $\Xi_-$}}
\put(64,114){\vector(1,0){14}}
\put(78,106){\vector(-1,0){14}}
\put(72.5,118){\makebox(0,0){\scriptsize $A^{-}$}}
\put(72.5,102){\makebox(0,0){\scriptsize $A^{+}$}}
\put(100,86){\vector(1,0){14}}
\put(107.5,90){\makebox(0,0){\scriptsize $A^{-}$}}
\put(100,80){\vector(0,-1){14}}
\put(96,73){\makebox(0,0){\scriptsize $\Sigma_-$}}
\put(65,170){\makebox(0,0){\scriptsize $F^{(2)}_{n\:1}$}}
\put(80,170){\makebox(0,0){\scriptsize $F^{(1)}_{n\:1}$}}
\put(100,170){\makebox(0,0){\scriptsize $F^{(2)}_{n\:2}$}}
\put(115,170){\makebox(0,0){\scriptsize $F^{(1)}_{n\:2}$}}
\put(135,170){\makebox(0,0){\scriptsize $F^{(2)}_{n\:3}$}}
\put(150,170){\makebox(0,0){\scriptsize $F^{(1)}_{n\:3}$}}
\put(170,170){\makebox(0,0){\scriptsize $F^{(2)}_{n\:4}$}}
\put(185,170){\makebox(0,0){\scriptsize $F^{(1)}_{n\:4}$}}
\put(205,170){\makebox(0,0){\scriptsize $F^{(2)}_{n\:5}$}}
\put(220,170){\makebox(0,0){\scriptsize $F^{(1)}_{n\:5}$}}
\linethickness{0.2mm}
\multiput(60,12)(35,0){5}{\line(1,0){25}}
\multiput(60,11)(35,0){5}{\line(0,1){2}}
\multiput(85,11)(35,0){5}{\line(0,1){2}}
\multiput(60,165)(35,0){5}{\line(1,0){25}}
\multiput(60,164)(35,0){5}{\line(0,1){2}}
\multiput(85,164)(35,0){5}{\line(0,1){2}}
\linethickness{0.3mm}
\multiput(75,20)(2,0){5}{\line(1,0){1}}
\multiput(110,83)(2,0){5}{\line(1,0){1}}
\multiput(145,116)(2,0){5}{\line(1,0){1}}
\multiput(180,139)(2,0){5}{\line(1,0){1}}
\multiput(215,157)(2,0){5}{\line(1,0){1}}
\end{picture}
\caption{\footnotesize {Energy levels for the relativistic Kepler-Coulomb problem. The dashed lines correspond to the null component of the Schr\"odinger ground state and the radial quantum number $n=N-|k|$, with $N=1,2,3,...$ the principal quantum number. The action of the operators $A^{\pm}$, $\Sigma_\pm$ and $\Xi_\pm$ on the radial states is shown.}}
\end{figure}
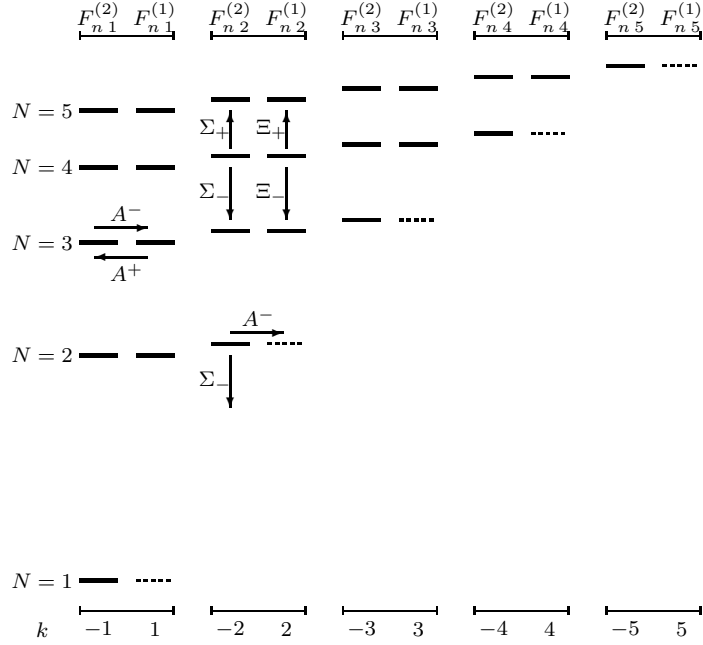

From the definition (\ref{n}), equations (\ref{Sigma+-}) and (\ref{Xi+-}) imply
\begin{align}
\Sigma_\pm\psi_s^n& \propto\psi_s^{n\pm1},\label{gs1}\\
\Xi_\pm\psi_{s+1}^{n-1}& \propto \psi_{s+1}^{n-1\pm1}.\label{gs2}
\end{align}

It must be notice that equation (\ref{gs2}) is valid for any value of the radial quantum number except for $n=0$. This is because the upper component of the spinor for $n=1$ can not be obtained from the action of the operator $\Xi_+$ on the upper component of the Schr\"odinger ground state (equation (\ref{spinor2})). These results are shown in Figure 1.
We illustrate the action of the SUSY operators $A^{\pm}$ on the eigenstates of the relativistic Kepler-Coulomb problem \cite{SUKU}. Also, it is shown that the lower components for the states corresponding to $n=0$ are annihilated by the operators $\Sigma_-$ and $A^{-}$. This result and the fact that the upper component of the SUSY \cite{SUKU} and the Schr\"odinger ground states are zero imply the equality of these states.

Equations (\ref{T3dif2or1}), (\ref{gs1}) and (\ref{gs2}) 
allow us to show that the action of the Schr\"odinger operators on the
states $\psi_s^{n}$ is
\begin{align}
B_\pm\psi_s^{n}& \propto\psi_s^{n\pm1}.\label{s1}
\end{align}
A similar result can be obtained from the equation (\ref{n}) and
the operators $\Xi_\pm$. These results
imply that the action of the Schr\"odinger operators on the
components of the spinor $\Phi_{k}^n$ is to change only the radial
quantum number $n$ leaving fixed the Dirac quantum number $k=k(s)$.

\section{Concluding Remarks}
We have shown that the algebraic treatment for the radial
equations of the relativistic Kepler-Coulomb problem is reduced to
find the non-compact symmetries for equation (\ref{desajun1}) or
(\ref{desajun2}). By applying the Schr\"odinger factorization we
found the one-variable generators for the $su(1,1)$ Lie algebra.
From the theory of unitary representations and the relation between
the components of the spinor $\Phi_k^n$ we found the energy spectrum
for this system in a purely algebraic way.
Moreover, equations (\ref{gs1}) and (\ref{gs2}) imply that the
$su(1,1)$ algebra generators $\Xi_{\pm}$, $\Xi_3$, and
$\Sigma_{\pm}$, $\Sigma_3$, are represented by
infinite-dimensional Hilbert subspaces of the radial quantum
states. The non-compact nature of the $su(1,1)$ algebra for this
problem reflects that for a fixed Dirac quantum number, the radial
quantum number is bounded from below and unbounded from above. We showed that
the Schr\"odinger ground state, which corresponds to the lowest value of the radial quantum number $n=0$, 
is equal to the SUSY ground state \cite{SUKU,DAHL}.
Finally, we emphasize that our treatment does not introduce an
extra variable and shows the origin of the $su(1,1)$ Lie algebra
generators.

\section*{Acknowledgments}
This work was partially supported by SNI-M\'exico, CONACYT grant
number J1-60621-I, COFAA-IPN, EDI-IPN, SIP-IPN projects numbers
 20100897  and 20100518, and ADI-UACM project number 7DA2023001.

\end{document}